\documentclass[twoside,version1996,preprint,9pt]{siggraph-99}

\usepackage{epsfig}
\usepackage{times}
\usepackage{mathptm}
\usepackage{amsfonts}
\usepackage{graphicx}
\usepackage{fancyhdr}

\newlength{\headrulelength}
\newlength{\headrulegap}
\def\headrule{\hspace{\headrulegap}\rule[2ex]{\headrulelength}{\headrulewidth}\gdef\headrule{\hrule}}

\fancypagestyle{empty}{
  \fancyfoot{}
  \fancyhead{}
  \fancyhead[RO,LE]{\fontsize{8}{8}\textsf{Computer Graphics Proceedings, Annual Conference Series, 1999}}
  \fancyfoot[RO,LE]{\thepage}
  \fancyhead[LO,RE]{\includegraphics[height=.66in]{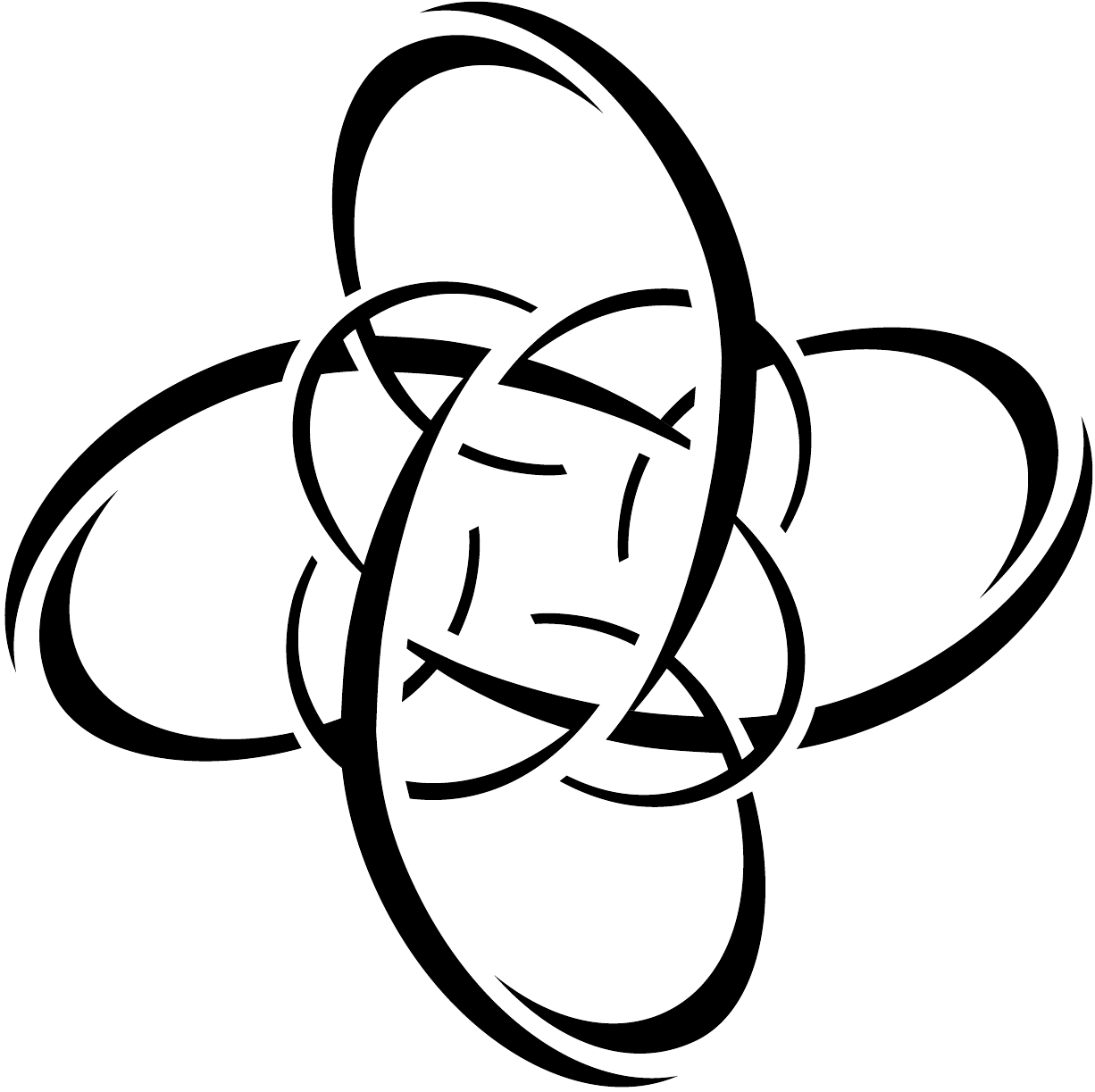}\vspace*{-.46in}}
}

\title{Shape Transformation Using Variational Implicit Functions}
\author{Greg Turk %
\and James F. O'Brien}
\affiliation{Georgia Institute of Technology}

\acmcategory{research}
\acmformat{print}

\contactname{Greg Turk}
\contactaddress{College of Computing \\
   Georgia Institute of Technology \\
   801 Atlantic Drive
   Atlanta, GA 30332-0280}
\contactphone{(404) 894--7508}
\contactfax{(404) 894--0673}
\contactemail{turk@cc.gatech.edu}

\estpages{8}

\suppresscover

\begin{document}

\maketitle

\baselineskip = .95\baselineskip

\begin{abstract}

Traditionally, shape transformation using implicit functions is performed in
two distinct steps: 1) creating two implicit functions, and 2) interpolating
between these two functions.  We present a new shape transformation method
that combines these two tasks into a single step.  We create a
transformation between two $N$-dimensional objects by casting this as a
scattered data interpolation problem in $N+1$ dimensions.  For the case of 2D
shapes, we place all of our data constraints within two planes, one for each
shape.  These planes are placed parallel to one another in 3D.  Zero-valued
constraints specify the locations of shape boundaries and positive-valued
constraints are placed along the normal direction in towards the center of
the shape.  We then invoke a variational interpolation technique (the 3D
generalization of thin-plate interpolation), and this yields a single
implicit function in 3D.  Intermediate shapes are simply the zero-valued
contours of 2D slices through this 3D function.  Shape transformation
between 3D shapes can be performed similarly by solving a 4D interpolation
problem.  To our knowledge, ours is the first shape transformation method to
unify the tasks of implicit function creation and interpolation.  The
transformations produced by this method appear smooth and natural, even
between objects of differing topologies.  If desired, one or more additional
shapes may be introduced that influence the intermediate shapes in a
sequence.  Our method can also reconstruct surfaces from multiple slices
that are not restricted to being parallel to one another.

\end{abstract}


\begin{CRcatlist}
    \CRcat{I.3.5}{Computer Graphics}%
{Computational Geometry and Object Modeling}%
{surfaces and object representations}
\end{CRcatlist}

\keywordlist

\ifcameraelse{
  \renewcommand{\thefootnote}{}%
  \footnotetext[0]{
    \vspace*{-0.10in}
    \par\noindent 
    turk@cc.gatech.edu, job@acm.org.\\
    \rule[1.3in]{0in}{0in}
  } 
  \renewcommand{\thefootnote}{\arabic{footnote}}
  \preprinttext{}  
}{
  \renewcommand{\thefootnote}{}%
  \footnotetext[0]{
    \par\noindent 
     turk@cc.gatech.edu, job@acm.org.\\
    \parbox[t][1in][b]{\columnwidth}{
      {\large \bf Authors' pre-print version.}\\
      \textbf{SIGGRAPH 99, Los Angeles, CA USA}
    }
  }
  \preprinttext{}  
  \renewcommand{\thefootnote}{\arabic{footnote}}
}


\begin{figure}[!t]
\centerline{\epsfig{file=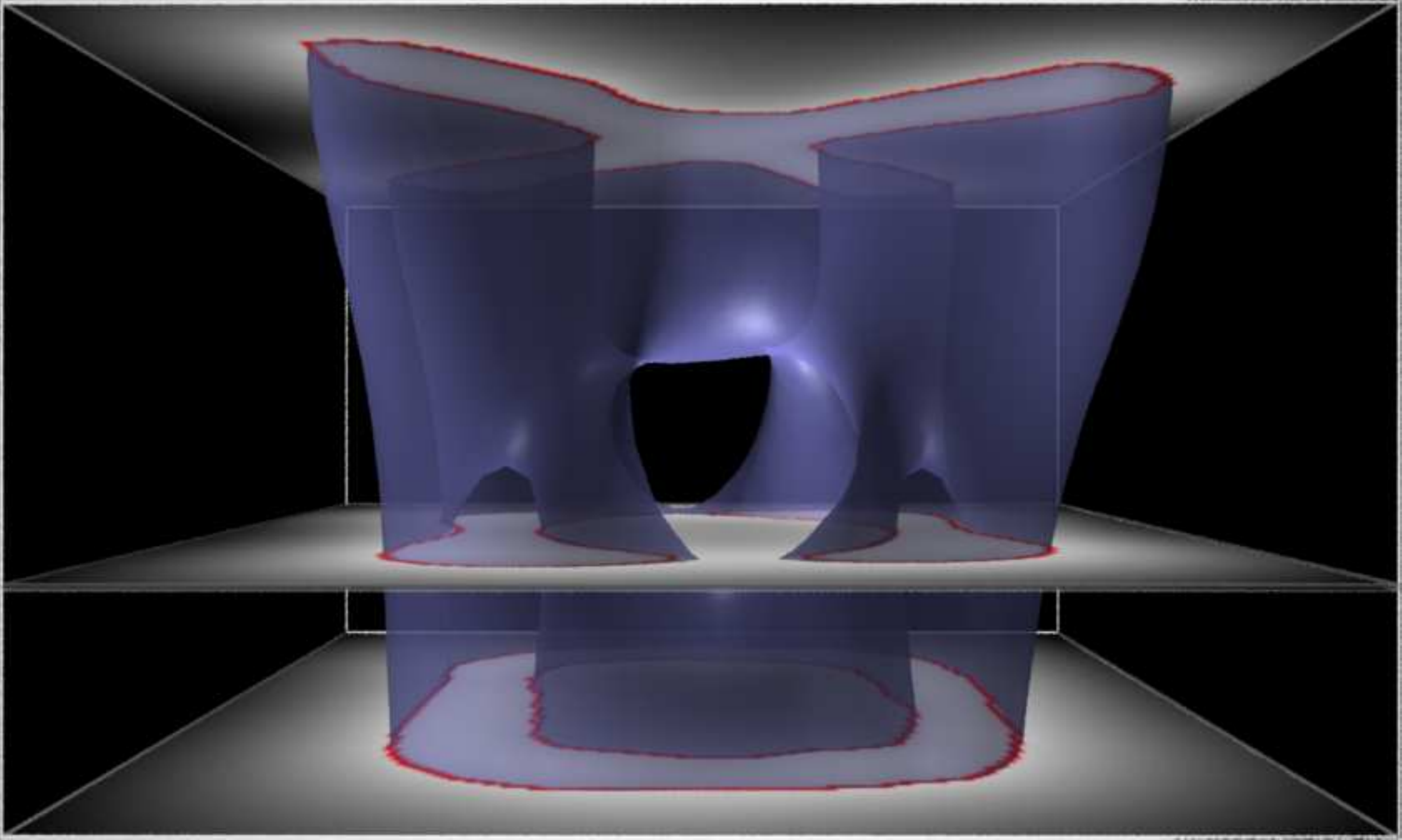,width=3.3in}}
\vspace*{-0.10in}
\caption{
  Visualization of transformation between X and O shapes.  Top and bottom
  planes contain constraints for the two shapes.  Translucent surface
  is the isosurface of a 3D variational implicit function, and slices
  through it give intermediate shapes.
}
\label{fig:surface_slice}
\vspace*{-0.15in}
\end{figure}

\section{Introduction}

The shape transformation problem can be stated as follows:  Given two shapes
A and B, construct a sequence of intermediate shapes so that adjacent pairs
in the sequence are geometrically close to one another.  Playing the
resulting sequence of shapes as an animation would show object A deforming
into object B.  Sequences of 2D shapes can be thought of as slices through a
3D surface, as shown in Figure~\ref{fig:surface_slice}.  Shape
transformation can be performed between objects of any dimension, although
2D and 3D shapes are by far the most common cases.  Shape transformation has
applications in medicine, computer aided design, and special effects
creation.  We give an overview of these three applications below.

One important application of shape transformation in medicine is contour
interpolation.  Non-invasive imaging techniques often collect data about a
patient's internal anatomy in ``slices'' of a particular size such as $512
\times 512$ samples.  Usually many fewer slices are taken along the third
dimension so that a resulting volume might, for example, be sampled at $512
\times 512 \times 30$ resolution.  To reconstruct a 3D model of a particular
organ, the samples are segmented to create shapes (contours) within the
slices.  Intermediate shapes are then created between slices in the sparsely
sampled dimension.  The reconstructed 3D object is formed by stacking
together the original and the interpolated contours.  This is an example of
2D shape transformation.

Shape transformation can also be a useful tool in computer aided geometric
design.  Consider the problem of creating a join between two metal parts
with different cross-sections.  It is important for the connecting surface
to be smooth because those places with sharp ridges or creases are the
locations that are most likely to form cracks.  The intermediate surface
joining the two parts can be created using shape transformation, much in the
same way as with contour interpolation for medical imaging.  Because of the
smoothness properties of variational interpolation methods, we consider them
a natural tool to explore for shape transformation in CAD.

Finally, animated shape transformations have been used to create dramatic
special effects for feature films and commercials.  One of the best-known
examples of shape transformation is in the film \emph{Terminator 2}.  In
this film, a cyborg policeman undergoes a number of transformations from an
amorphous and highly reflective surface to various destination shapes.  2D
image morphing would not have accurately modeled the reflection of the
environment off the surface of the deforming cyborg, hence tailor-made 3D
shape transformation programs were used for these effects \cite{Duncan91}.

In this paper we use variational interpolation in a new way to produce
high-quality shape transformations that may be used for any of the
previously mentioned applications.  Our method allows a user to control the
transformation in several ways, and it is general enough to produce
transformations between shapes of any topology.

\vspace*{-0.15in}

\section{Previous Work}
\label{sec:previous_work}

Most shape transformation techniques can be placed into one of two
categories: parametric correspondence methods and implicit function
interpolation.  Parametric methods are typically faster to compute and
require less memory because they operate on a lower-dimensional
representation of an object than do implicit function methods.
Unfortunately, transforming between objects of different topologies is
considerably more difficult with parametric methods.  Parametric approaches
also can suffer from problems with self-intersecting surfaces, but this is
never a problem with implicit function methods.  Techniques that use
implicit function interpolation gracefully handle changes in topology
between objects and do not create self-intersecting surfaces.

A parametric correspondence approach to shape transformation attempts to
find a ``reasonable'' correspondence between pairs of locations on the
boundaries of the two shapes.  Intermediate shapes are then created by
computing interpolated positions between the corresponding pairs of points.
Many shape transformation techniques have been created that follow the
parametric correspondence approach.  One early application of this approach
is the method of contour interpolation described by Fuchs, Kedem and Uselton
\cite{Fuchs77}.  Their method attempts to find an ``optimal'' (minimum-area)
triangular tiling that connects two contours using dynamic programming.
Many subsequent techniques followed this approach of defining a quality
measure for a particular correspondence between contours and then invoking
an optimization procedure \cite{Meyers_Skinner91, Sederberg_Greenwood92}.
There have been fewer examples of using parametric correspondence for
3D shape transformation.  One quite successful 3D parametric method is the
work of Kent et al.~\cite{Kent92}.  The key to their approach is to subdivide
the polygons of the two models in a manner that creates a correspondence
between the vertices of the two models.  More recently, Gregory and co-workers
created a similar method that also allows a user to specify region
correspondences between meshes to better control a
transformation~\cite{Gregory98}.

An entirely different approach to shape transformation is to create an
implicit function for each shape and then to smoothly interpolate between
these two functions.  A shape is defined by an implicit function, $f({\bf
x})$, as the set of all points ${\bf x}$ such that $f({\bf x}) = 0$.  For
contour interpolation in 2D, the implicit function can be thought of as a
height field over a two-dimensional domain, and the boundary of a shape is
the one-dimensional curve defined by all the points that have the same
elevation value of zero.  An implicit function in 3D is a function that
yields a scalar value at every point in 3D.  The shape described by such a
function is given by those places in 3D whose function value is zero (the
isosurface).  

One commonly used implicit function is the \emph{inside/outside function} or
\emph{characteristic function}.  This function takes on only two values over
the entire domain.  The two values that are typically used are zero to
represent locations that are outside and one to signify positions that are
inside the given shape.  Given a powerful enough interpolation technique,
the characteristic function can be used for creating shape transformations.
Hughes presented a successful example of this approach by transforming
characteristic functions into the frequency domain and performing
interpolation on the frequency representations of the shapes
\cite{Hughes92}.  Kaul and Rossignac found that smooth intermediate shapes
can be generated by using weighted Minkowski sums to interpolate between
characteristic functions \cite{Kaul_Rossignac91}.  They later created a
generalization of this technique that can use several intermediate
shapes to control the interpolation between objects \cite{Rossignac_Kaul94}.
Using a wavelet decomposition of a characteristic function allowed He and
colleagues to create intermediates between quite complex 3D objects
\cite{He94}.

A more informative implicit function can provide excellent intermediate
shapes even if a simple interpolation technique is used.  In particular, the
\emph{signed distance function} (sometimes called the \emph{distance
transform}) is an implicit function that gives very plausible intermediate
shapes even when used with simple linear interpolation of the function
values of the two shapes.  The value of the signed distance function at a
point ${\bf x}$ inside a given shape is just the Euclidean distance between
${\bf x}$ and the nearest point on the boundary of the shape.  For a point
${\bf x}$ that is outside the shape, the signed distance function takes on
the negative of the distance from ${\bf x}$ to the closest point on the
boundary.

Several researchers have used the signed distance function to interpolate 
between 2D contours \cite{Levin86, Herman92}.  The distance function 
for each given shape is represented as a regular 2D grid of values, and an 
intermediate implicit function is created by linear interpolation between 
corresponding grid values of the two implicit functions.  Each intermediate 
shape is given by the zero iso-contour of this interpolated implicit function.  
In contrast to the global interpolation methods described above (frequency 
domain, wavelets, Minkowski sum), this interpolation is entirely local in 
nature.  Nevertheless, the shape transformations that are created by this 
method are quite good.  In essence, the information that the signed distance 
function encodes (distance to nearest boundary) is enough to make up for 
the purely local method of interpolation.  Payne and Toga were the first to 
transform three dimensional shapes using this approach~\cite{Payne_Toga92}.
Cohen-Or and colleagues gave additional control to this same approach by
combining it with a warping technique in order to produce shape
transformations of 3D objects~\cite{Cohen-Or97}.

Our approach to shape transformation combines the two steps of building
implicit functions and interpolating between them.  To our knowledge, it is
the only method to do so.  The remainder of this paper describes how
variational interpolation can be used to simultaneously solve these
two tasks.

\section{Variational Interpolation}
\label{sec:variational_interpolation}

Our approach relies on \emph{scattered data interpolation} to solve the
shape transformation problem.  The problem of scattered interpolation is to
create a smooth function that passes through a given set of data points.
The two-dimensional version of this problem can be stated as follows:  Given
a collection of $k$ constraint points $\{{\bf c}_1, {\bf c}_2,\ldots,{\bf
c}_k\}$ that are scattered in the plane, together with scalar height values
at each of these points $\{h_1, h_2,\ldots, h_k\}$, construct a smooth
surface that matches each of these heights at the given locations.  We can
think of this solution surface as a scalar-valued function $f({\bf x})$ so
that $f({\bf c}_i) = h_i$, for $1 \leq i \leq k$.

One common approach to solving scattered data problems is to use variational
techniques (from the calculus of variations).  This approach begins with an
energy that measures the quality of an interpolating function and then finds
the single function that matches the given data points and that minimizes
this energy measure.  For two-dimensional problems, thin-plate interpolation
is the variational solution when using the following energy function $E$:

\begin{equation}
  E = \int_{\Omega}{f_{xx}^{2}({\bf x}) + 2 f_{xy}^{2}({\bf x})
    + f_{yy}^{2}({\bf x})}
\label{eqn:energy}
\end{equation}

The notation $f_{xx}$ means the second partial derivative in the $x$
direction, and the other two terms are similar partial derivatives, one of
them mixed.  The above energy function is basically a measure of the
aggregate squared curvature of $f({\bf x})$ over the region of interest
$\Omega$.  Any creases or pinches in a surface will result in a larger value
of $E$.  A smooth surface that has no such regions of high curvature will
have a lower value of $E$.  The thin-plate solution to an interpolation
problem is the function $f({\bf x})$ that satisfies all of the constraints
and that has the smallest possible value of $E$.

\begin{figure}[!t]
\centerline{\epsfig{file=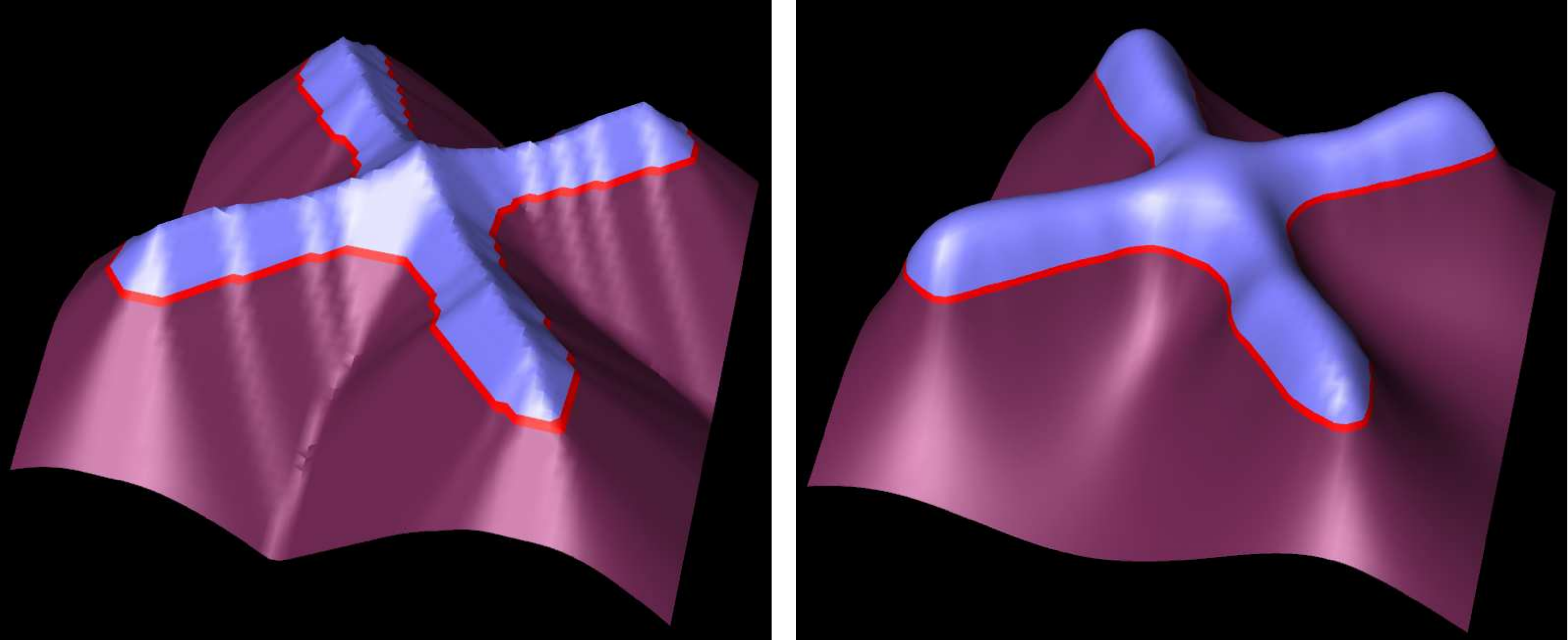,width=3.5in}}
\caption{
  Implicit functions for an X shape.  Left shows the signed distance
  function, and right shows the smoother variational implicit function.
}
\vspace*{-0.15in}
\label{fig:x_axes}
\end{figure}

The scattered data interpolation problem can be formulated in any number of
dimensions.  When the given points ${\bf c}_i$ are positions in $N$-dimensions
rather than in 2D, this is called the $N$-dimensional scattered data
interpolation problem.  There are appropriate generalizations to the energy
function and to thin-plate interpolation for other dimensions.  In this paper
we will perform interpolation in two, three, four and five dimensions.
Because the term \emph{thin-plate} is only meaningful for 2D problems, we
will use \emph{variational interpolation} to mean the generalization of
thin-plate techniques to any number of dimensions.

The scattered data interpolation task as formulated above is a variational
problem where the desired solution is a function, $f({\bf x})$, that will
minimize equation~\ref{eqn:energy} subject to the interpolation constraints
$f({\bf c}_i) = h_i$.  Equation~\ref{eqn:energy} can be solved using
weighted sums of the radial basis function $\phi ({\bf x}) = {|{\bf x}|}^2
\log (|{\bf x}|)$.  The family of variational problems that includes
equation~\ref{eqn:energy} was studied by Duchon~\cite{Duchon77}.

Using the appropriate radial basis function, we can then express the 
interpolation function as

\begin{equation}
  f({\bf x}) = \sum_{j=1}^{n} d_j \phi ({\bf x}-{\bf c}_j) + P({\bf x})
\label{eqn:sum_of_basis}
\end{equation}

In the above equation, ${\bf c}_j$ are the locations of the constraints, the
$d_j$ are the weights, and $P({\bf x})$ is a degree one polynomial that
accounts for the linear and constant portions of $f$.  Because the
thin-plate radial basis function naturally minimizes
equation~\ref{eqn:energy}, determining the weights, $d_j$, and the
coefficients of $P({\bf x})$ so that the interpolation constraints are
satisfied will yield the desired solution that minimizes
equation~\ref{eqn:energy} subject to the constraints.  Furthermore, the
solution will be an exact analytic solution, and is not subject to
approximation and discretization errors that may occur when using finite
element or finite difference methods.

To solve for the set of $d_j$ that will satisfy the interpolation
constraints $h_i = f({\bf c}_i)$, we can substitute the right side of
equation~\ref{eqn:sum_of_basis} for $f({\bf c}_i)$, which gives:

\begin{eqnarray}
 h_i &=& \sum_{j=1}^k d_j \phi({\bf c_i} - {\bf c_j}) + P({\bf c_i})
\end{eqnarray}

Since this equation is linear with respect to the unknowns, $d_j$ and the
coefficients of $P({\bf x})$, it can be formulated as a linear system.  For
interpolation in 3D, let ${\bf c}_i = (c_i^x, c_i^y, c_i^z)$ and let
$\phi_{ij} = \phi({\bf c}_i-{\bf c}_j)$.  Then this linear system can be
written as follows:

\vspace*{-0.15in}

\begin{displaymath}
\left[ 
 \begin{array}{ccccccccc}
\phi_{11} & \phi_{12} & \dots & \phi_{1k} & 1     & c_1^x  & c_1^y  & c_1^z \\
\phi_{21} & \phi_{22} & \dots & \phi_{2k} & 1     & c_2^x  & c_2^y  & c_2^z \\
\vdots    &  \vdots   &       & \vdots   & \vdots & \vdots & \vdots & \vdots \\
\phi_{k1} & \phi_{k2} & \dots & \phi_{kk} & 1     & c_k^x  & c_k^y  & c_k^z \\
1         & 1         & \dots & 1         & 0     & 0      & 0      & 0  \\
c_1^x     & c_2^x     & \dots & c_k^x     & 0     & 0      & 0      & 0  \\
c_1^y     & c_2^y     & \dots & c_k^y     & 0     & 0      & 0      & 0  \\
c_1^z     & c_2^z     & \dots & c_k^z     & 0     & 0      & 0      & 0       
 \end{array}
\right] 
\left[ 
  \begin{array}{c}
    d_1    \\
    d_2    \\
    \vdots \\
    d_k    \\
    p_0    \\
    p_1    \\
    p_2    \\
    p_3    
  \end{array}
\right] 
=
\left[ 
  \begin{array}{c}
    h_1    \\
    h_2    \\
    \vdots \\
    h_k    \\
    0      \\
    0      \\
    0      \\
    0     
  \end{array}
\right] 
 \begin{array}{c}
 \\ \\ \\ \\ \\ \\ \\ \\ \\ \\ \\ \\
  \end{array}
\vspace*{-0.15in}
\end{displaymath}  

The above system is symmetric and positive semi-definite, so there will
always be a unique solution for the $d_j$ and $p_j$ \cite{Golub_vanLoan96}.
For systems with up to a few thousand constraints, the system can be solved
directly with a technique such as symmetric LU decomposition.  We used
symmetric LU decomposition to solve this system for all of the examples
shown in this paper.

Using the tools of variational interpolation we can now turn our attention
to creating implicit functions for shape transformation.

\section{Smooth Implicit Function Creation}
\label{sec:function_creation}

In this section we will lay down the groundwork for shape transformation by
discussing the creation of smooth implicit functions for a single shape.  In
particular, we will use variational interpolation of scattered constraints
to construct implicit functions.  Later we will generalize this to create
functions that perform shape transformation.

\begin{figure}[!t]
\centerline{\epsfig{file=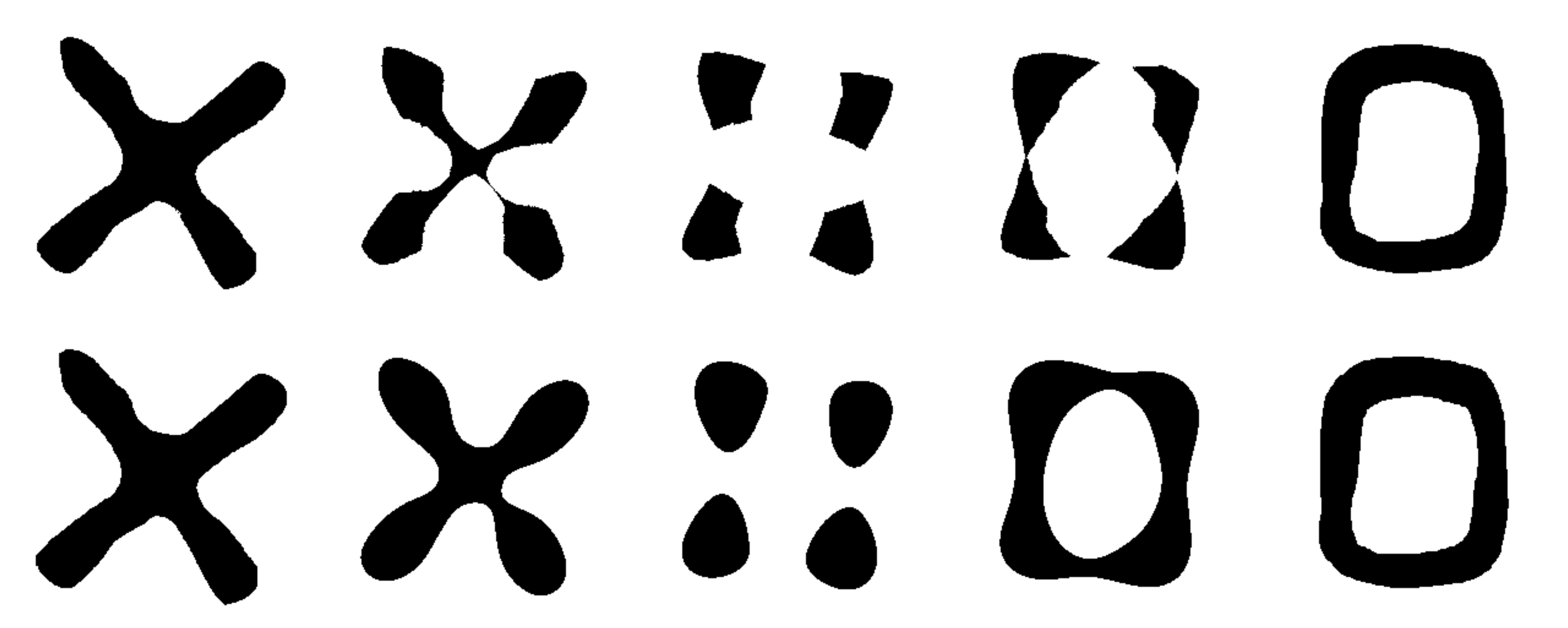,width=3.8in}}
\caption{
  Upper row is a shape transformation created using the signed distance
  transform.  Lower row is the sequence generated using a single variational
  implicit function.
}
\vspace*{-0.15in}
\label{fig:xo_sequences}
\end{figure}

Let us first examine the signed distance transformation because it is
commonly used for shape transformation.  The left half of
Figure~\ref{fig:x_axes} shows a height field representation of the signed
distance function of an X shape.  The figure shows sharp ridges (the
\emph{medial axis}) that run down the middle of the height field.  Ridges
appear in the middle of shapes where the points are equally distant from two
or more boundary points of the original shape.  The values of a signed
distance function decrease as one moves away from the ridge towards the
boundaries.  Figure~\ref{fig:xo_sequences}, top row, shows a shape
interpolation sequence between an X and an O shape that was created by
linear interpolation between two signed distance functions.  Note the
pinched portions of some of the intermediate shapes.  These sharp features
are not isolated problems, but instead persist over many intermediate
shapes.  The cause of these pinches are the sharp ridges of signed distance
functions.  In many applications such artifacts are undesirable.  In medical
reconstruction, for example, these pinches are a poor estimate of shape
because most biological structures have smooth surfaces.  Because of this,
we seek implicit functions that are continuous and that have a continuous
first derivative.

\subsection{Variational Implicit Functions in 2D}

We can create smooth implicit functions for a given shape using variational
interpolation.  This can be done both for 2D and 3D shapes, although we will
begin by discussing the 2D case.  In this approach, we create a closed 2D
curve by describing a number of locations through which the curve will pass
and also specifying a number of points that should be interior to the curve.
We call the given points on the curve the \emph{boundary constraints}.  The
boundary constraints are locations at which we require our implicit function
to take on the value of zero.  Paired with each boundary constraint is a
\emph{normal constraint}, which is a location at which the implicit function
is required to take on the value one.  (Actually, any positive value could
be used.)  The locations of the normal constraints should be towards the
interior of the desired curve, and also the line passing through the normal
constraint and its paired boundary constraint should be parallel to the
desired normal to the curve.  The collection of boundary and normal
constraints are passed along to a variational interpolation routine as the
scattered constraints to be interpolated.  The function that is returned is
an implicit function that describes our curve.  The curve will exactly pass
through our boundary constraints.

Figure~\ref{fig:constraints_2d} (left) illustrates eight such pairs of
constraints in the plane, with the boundary constraints shown as circles and
the normal constraints as plus signs.  When we invoke variational
interpolation with such constraints, the result is a function that takes on
the value of zero exactly at our zero-value constraints and that is positive
in the direction of our normal constraints (towards the interior of the
shape).  The closed curve passing through the zero-value constraints in
Figure~\ref{fig:constraints_2d} (middle) is the iso-contour of the implicit
function created by this method.  Figure~\ref{fig:constraints_2d} (right)
shows the resulting implicit function rendered as a height field.  Given
enough suitably-placed boundary constraints we can define any closed shape.
We call an implicit function that is created in this manner a
\emph{variational implicit function}.  This new technique for creating
implicit functions also show promise for surface modeling, a topic that is
explored in \cite{Turk99}.

\begin{figure}[!t]
\centerline{\epsfig{file=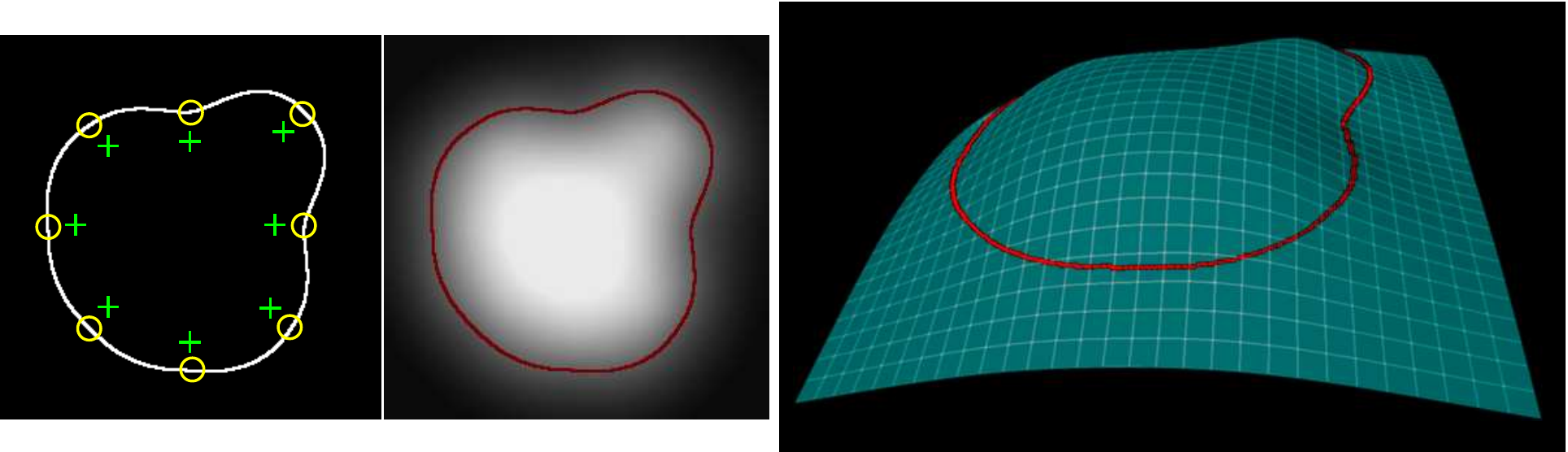,width=3.8in}}
\caption{
  At left are pairs of boundary and normal constraints (circles and pluses).
  The middle image uses intensity to show the resulting variational implicit
  function, and the right image shows the function as a height field.
}
\vspace*{-0.15in}
\label{fig:constraints_2d}
\end{figure}

We now turn our attention to defining boundary and normal constraints for a
given 2D shape.  Assume that a given shape is represented as a gray-scale
image.  White pixels represent the interior of a shape, black pixels will be
outside the shape, and pixels with intermediate gray values lie on the
boundary of the shape.  Let $m$ be the middle gray value of our image's gray
scale range.  Our goal is to create constraints between any two adjacent
pixels where one pixel's value is less than $m$ and the other's value is
greater.  Identifying these locations is the 2D analog of finding the vertex
locations in a 3D marching cubes algorithm \cite{Lorenson_Cline87}.

We traverse the entire gray-scale image and examine the east and south
neighbor of each pixel $I(x,y)$.  If $I(x,y) < m$ and either neighbor has a
value greater than $m$, we create a boundary constraint at a point along the
line segment joining the pixel centers.  A boundary constraint is also
created if $ I(x,y) > m$ and either neighbor takes on a value less than $m$.
The value of the constraint is zero, and we set the position of the
constraint at the location between the two pixels where the image would take
on the value of $m$ if we assume linear interpolation of pixel values.
Next, we estimate the gradient of the gray scale image using linear
interpolation of pixel values and central differencing.  We then create a
normal constraint a short distance away from the zero crossing in the
direction of the gradient.  We have found that a distance of a pixel's width
between the boundary and normal constraints works well in practice.
Figure~\ref{fig:x_axes} (right) shows the implicit function for an X shape
that was created using variational interpolation from such constraints.  It
is smooth and free of sharp ridges.

\subsection{Variational Implicit Functions in 3D}

We can create implicit functions for 3D surfaces using variational
interpolation in much the same way as for 2D shapes.  Specifically, we can
derive 3D constraints from the vertex positions and surface normals of a
polygonal representation of an object.  Let $(x,y,z)$ and $(n_x,n_y,n_z)$ be
the position and the surface normal at a vertex, respectively.  Then a
boundary constraint is placed at $(x,y,z)$ and a normal constraint is placed
at $(x-k n_x, y-k n_y, z-k n_z)$, where $k$ is some small value.  We use a
value of $k = 0.01$ for models that fit within a unit cube for the results
shown in this paper.  All of the 3D models that we transform
in this paper were constructed by building an implicit function in this
manner.  Note that we can use this method to build an implicit function
whenever we have a collection of points and normals--- polygon connectivity
is not necessary.

Now that we can construct smooth implicit functions for both two- and
three-dimensional shapes, we turn our attention to shape transformation.  It
would be possible to create variational implicit functions for each of two
given shapes and then linearly interpolate between these functions to create
a shape transformation sequence.  Instead, however, we will examine an
even better way of performing shape transformation by generalizing the
implicit function building methods of this section.

\begin{figure}[!b]
\vspace*{-0.15in}
\centerline{\epsfig{file=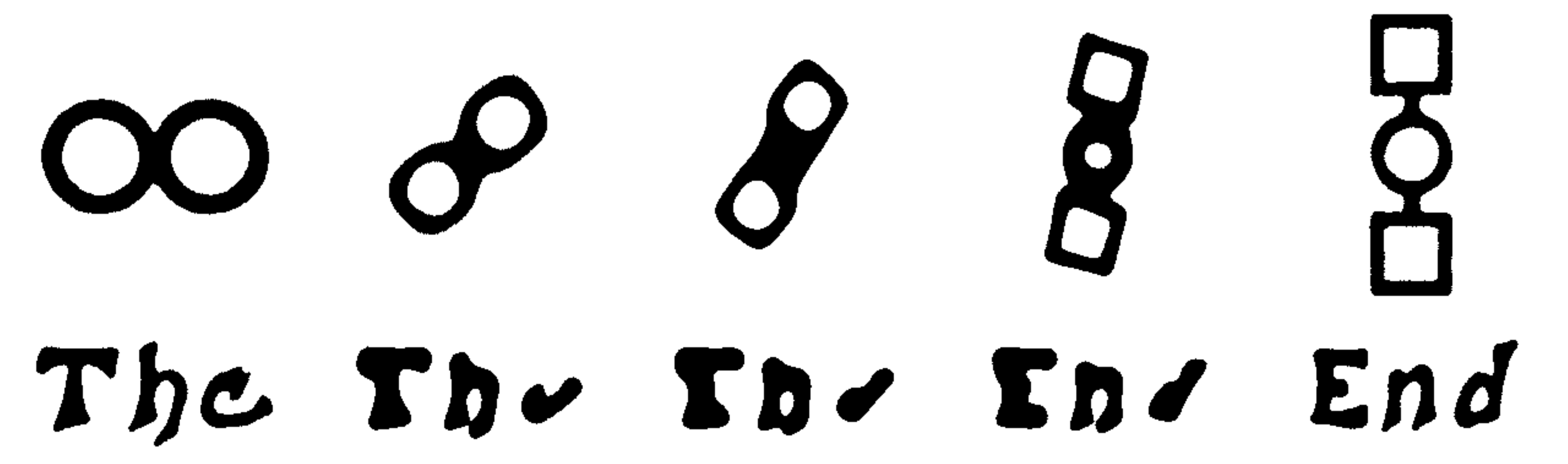,width=3.8in}}
\caption{
  Two shape transformation sequences (using the variational implicit
  technique) that incorporate warping.
}
\label{fig:inf_the_end}
\end{figure}

\section{Unifying Function Creation and Interpolation}
\label{sec:shape_interpolation}

The key to our shape transformation approach is to represent the entire
sequence of shapes with a single implicit function.  To do so, we need to
work in one higher dimension than the given shapes.  For 2D shapes, we will
construct an implicit function in 3D that represents our two given shapes in
two distinct parallel planes.  This is actually simple to achieve now that
we know how to use scattered data interpolation to create an implicit
function.

\subsection{Two-Dimensional Shape Transformation}

Given two shapes in the plane, assume that we have created a set of boundary
and normal constraints for each shape, as described in Section 4.  Instead
of using each set of constraints separately to create two different 2D
implicit functions, we will embed all of the constraints in 3D.  We do this
by adding a third coordinate value to the location of each boundary and
normal constraint.  For those constraints for the first shape, we set the
new coordinate $t$ for all constraints to $t=0$.  For the second shape, all
of the new coordinate values are set to $t=t_{max}$ (some non-zero value).
Although we have added a third dimension to the locations of the
constraints, the values that are to be interpolated remain unchanged for all
constraints.

Once we have placed the constraints of both shapes into 3D, we invoke 3D
variational interpolation to create a single scalar-valued function over
$\mathbf{R}^3$.  If we take a slice of this function in the plane $t=0$,
we find an implicit function that takes on the value zero exactly at
the boundary constraints for our first shape.  In this plane, our function
describes the first shape.  Similarly, in the plane $t=t_{max}$ this
function gives the second shape.  Parallel slices at locations between these
two planes ($0 < t < t_{max}$) represent the shapes of our shape
transformation sequence.  Figure~\ref{fig:surface_slice} illustrates that
the collection of intermediate shapes are all just slices through a surface
in 3D that is created by variational interpolation.

Figure~\ref{fig:xo_sequences} (bottom) shows the sequence of shapes created
using this variational approach to shape transformation.  Topology changes
(e.g. the addition or removal of holes) come ``for free'', without any human
guidance or algorithmic complications.  Notice that all of the intermediate
shapes have smooth boundaries, without pinches.  Sharp features can arise
only momentarily when there is a change in topology such as when two parts
join.  Figure~\ref{fig:inf_the_end} shows two more shape transformations
that use this approach and that also incorporate warping.  Warping is an
another degree of control that may be added to any shape transformation
technique, and is in fact an orthogonal issue to those of implicit function
creation and interpolation.  Although it is not a focus of our research, for
completeness we briefly describe warping in the appendix.

Why has this implicit function building method not been tried using other
ways of creating implicit functions?  Why not, for example, build a signed
distance function in one higher dimension?  Because a \emph{complete}
description of an object's boundary is required in order to build a signed
distance function.  When we embed our two shapes into a higher dimension, we
only know a \emph{piece} of the boundary of our desired higher-dimensional
shape, namely the cross-sections that match the two given objects.  In
contrast, a complete boundary representation is \emph{not} required when
using variational interpolation to create an implicit function.  Variational
interpolation creates plausible function values in regions where we have no
information, and especially in the ``unknown'' region between the two planes
that contain all of our constraints.  This plausibility of values comes from
the smooth nature of the functions that are created by the variational
approach.

\subsection{Three-Dimensional Shape Transformation}

Just as we create a 3D function to create a transformation between 2D
shapes, we can move to 4D in order to create a sequence between 3D shapes.
We perform shape interpolation between two 3D objects using boundary
and normal constraints for each shape.  We place the constraints from two 3D
objects into four dimensional space, just as we placed constraints from 2D
contours into 3D.  Similar to contour interpolation, the constraints are
separated from one another in the fourth dimension by some specified
distance.  We place all the constraints from the first object at $t=0$, and
the constraints from the second object are placed at $t=t_{max}$, where
$t_{max}$ is the given separation distance.  We then create a 4D implicit
function using variational interpolation.  An intermediate shape between the
two given shapes is found by extracting the isosurface of a 3D ``slice''
(actually a volume) of the resulting 4D function.

Figure~\ref{fig:bunstar} shows two 3D shape transformation sequence that
were constructed using this method.  To extract these surfaces we use code
published by Bloomenthal that begins at a seed location on the surface of a
model and only evaluates the implicit function at points near previously
visited locations \cite{Bloomenthal94}.  This is far more efficient than
sampling an entire volume of the implicit function and then extracting an
isosurface from the volume.  The matrix solution for the transformation
sequence of Figure~\ref{fig:bunstar} (left) required 13.5 minutes, and each
isosurface in the sequence took approximately 2.3 minutes to generate on an
SGI Indigo2 with a 195 MHz R10000 processor.  Figure~\ref{fig:bunstar}
(right) shows a transformation between 3D shapes that used warping to align
features.

\begin{figure}[!t]
\centerline{\epsfig{file=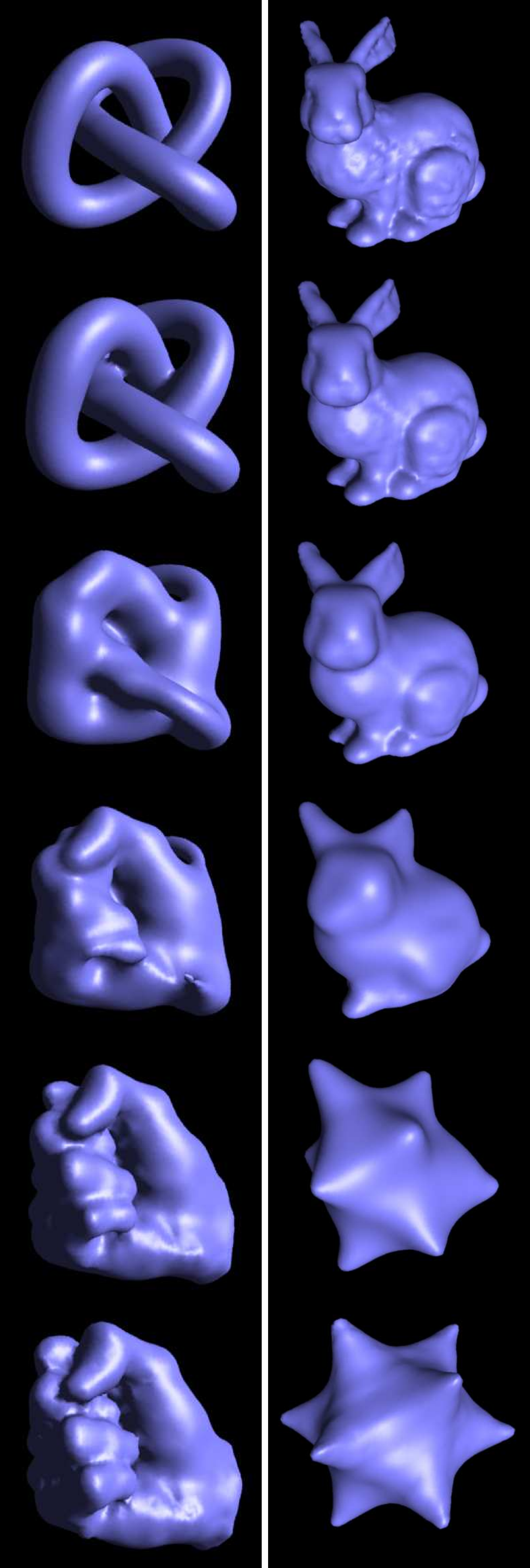,width=2.9in}}
\vspace*{-0.10in}
\caption{
 3D shape transformation sequences.
}
\vspace*{-0.25in}
\label{fig:bunstar}
\end{figure}



\begin{figure*}[!t]
\centerline{\epsfig{file=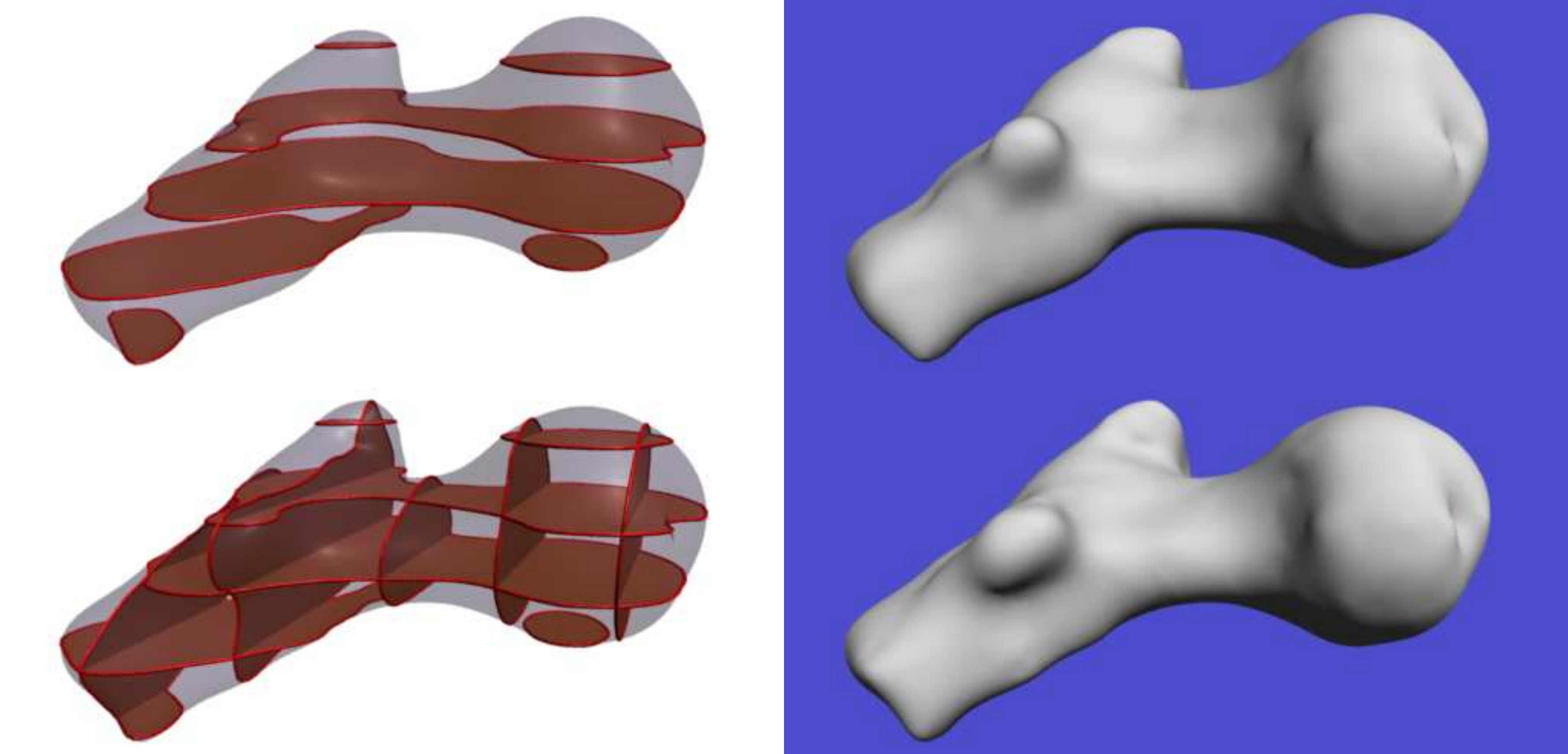,width=5.0in}}
\caption{
  Reconstruction of hip joint from contours.  Top row shows the five
  parallel slices used and the final surface.  Bottom row shows
  intersecting contours and the more detailed surface that is created.
}
\vspace*{-0.15in}
\label{fig:ball_joint}
\end{figure*}

\section{Surface Reconstruction from Contours}
\label{sec:contours}

So far we have only considered shape transformation between pairs of
objects.  In medical reconstruction, however, it is often necessary to
create a surface from a large number of parallel 2D slices.  Can't we just
perform shape interpolation between pairs of slices and stack the results
together to create one surface in 3D?  Although this method will create a
continuous surface, it is almost certain to produce a shape that has
surface normal discontinuities at the planes of the original slices.  In the
plane of slice $i$, the surface created between slice pairs $i-1$ and $i$
will usually not agree in surface normal with the surface created between
slices $i$ and $i+1$.  Nearly all contour interpolation methods consider
only pairs of contours at any one time, and thus suffer from such
discontinuities.  (A notable exception is \cite{Barequet96}).

To avoid discontinuities in surface normal, we must use information about
more than just two slices at a given time.  We can accomplish this using a
generalization of the variational approach to shape transformation.  Assume
that we begin with $k$ sets of constraints, one set for each 2D data slice.
Instead of considering the contours in pairs, we place the constraints for
all of the $k$ slices into 3D simultaneously.  Specifically, the constraints
of slice $i$ are placed in the plane $z = si$, where $s$ is the spacing
between planes.  Once the constraints from \emph{all} slices have been
placed in 3D, we invoke variational interpolation \emph{once} to create a
single implicit function in 3D.  The zero-valued isosurface exactly passes
through each contour of the data.  Due to the smooth nature of variational
interpolation, the gradient of the implicit function is everywhere
continuous.  This means that surface normal discontinuities are rare,
appearing in pathological situations when the gradient vanishes such as when
two features just barely touch.  Figure~\ref{fig:ball_joint} (top row)
illustrates the result of this contour interpolation approach.  The hip
joint reconstruction in the upper right was created from the five slices
shown at the upper left.

A side benefit of using the variational implicit function method is that it
produces smoothly rounded caps on the ends of surfaces.  Notice that in
Figure~\ref{fig:ball_joint} (top left) that the reconstructed surface
extends beyond the constraints in the positive and negative $z$ direction
(the direction of slice stacking).  This ``rounding off'' of the ends is a
natural side effect of variational interpolation, and need not be explicitly
specified.

\subsection{Non-Parallel Contours}

In the previous section, we only considered placing constraints within
planes that are all parallel to one another.  There is nothing special about
any particular set of planes, however, once we are specifying constraints in
3D.  We can mix together constraints that are taken from planes at any angle
whatsoever, so long as we know the relative positions of the planes (and
thus the constraints).  Most contour interpolation procedures cannot
integrate data taken from slices in several directions, but the variational
approach allows complete freedom in this regard.
Figure~\ref{fig:ball_joint} (lower row) shows several contours that are
placed perpendicular to one another, and the result of using variational
interpolation on the group of constraints from these contours.

\begin{figure*}[!t]
\centerline{\epsfig{file=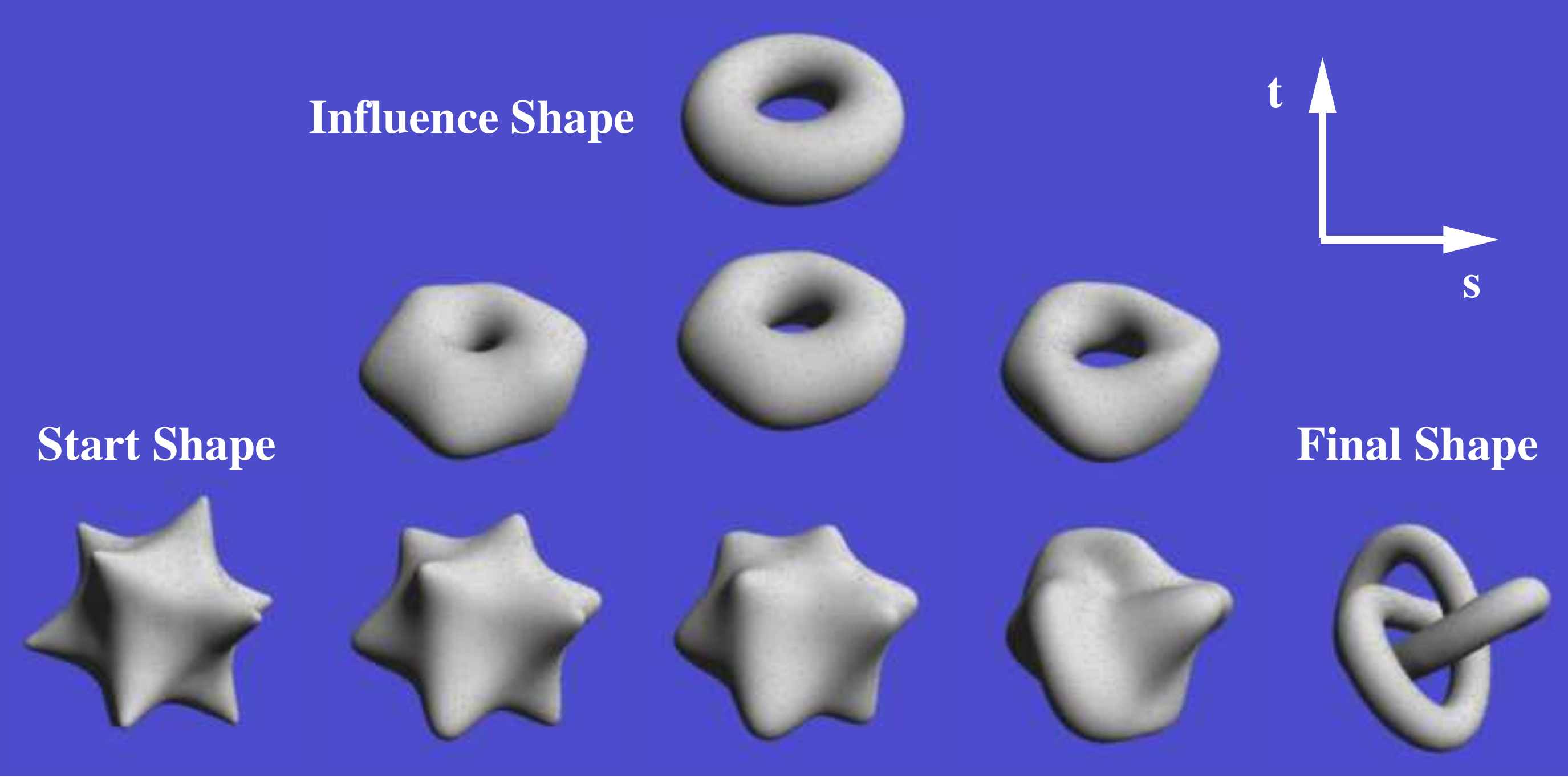,width=6.5in}}
\caption{
  Sequence between star and knot can be influenced by a torus (the influence
  shape) if the path passes near the torus in the five-dimensional space.
}
\vspace*{-0.15in}
\label{fig:influence}
\end{figure*}

\subsection{Between-Contour Spacing}

Up to this point we have not discussed the separating distance $s$
between the slices that contain the contour data.  This separating distance
has a concrete meaning in medical shape reconstruction from 2D contours.
Here we know the actual 3D separation between the contours from the data
capture process.  This ``natural'' distance is the separating distance $s$
that should be used when reconstructing the surface using variational
interpolation.  Upon reflection, it is odd that some contour interpolation
methods do not make use of the data capture distance between slices.  In
some cases a medical technician will deliberately vary the spacing between
data slices in order to capture more data in a particular region of
interest.  Using variational interpolation, we may incorporate this
information about varying separation distances into the surface
reconstruction process.

For both special effects production and for computer aided design, the
distance between the separating planes can be thought of as a control knob
for the artist or designer.  If the distance is small, only pairs of
features from the two shapes that are very close to one another will be
preserved through all the intermediate shapes.  If the separation distance
is large, the intermediate shape is guided by more global properties of the
two shapes.  In some sense, the separating distance specifies whether the
shape transformation is local or global in nature.  The separation distance
is just one control knob for the user, and in the next section we will
explore another user control.

\section{Influence Shapes}
\label{sec:influence_shapes}

In this section we present a method of controlling shape transformation by
introducing an \emph{influence shape}.  The idea to use additional objects
as controls for shape transformation was introduced by Rossignac and
Kaul \cite{Rossignac_Kaul94}.  Such intermediate shape control can be
performed in a natural way using variational interpolation.  The key is to
step into a still higher dimension when performing shape transformation.

Recall that to create a transformation sequence between two given shapes we
added one new dimension, called $t$ earlier.  We can think of the
two shapes as being two points that are separated along the $t$ dimension,
and these two points are connected by a line segment that joins the two
points along this dimension.  If we begin with three shapes, however, we can
in effect place them at the three points of a triangle.  In order to do so
we need not just one additional dimension but two, call them $s$ and $t$.

As an example, we may begin with three different 3D shapes A, B and C.  To
each constraint that describes one of the shapes, we can add two new
coordinates, $s$ and $t$.  Constraints from shape A at $(x,y,z)$ are placed
at $(x,y,z,0,0)$, constraints from shape B are placed at $(x,y,z,1,0)$ and
shape C constraints are placed at $(x,y,z,1/2,1/2)$.  Variational
interpolation based on these 5-dimensional constraints results in a 5D
implicit function.  Three-dimensional slices of this function along the
$s$-dimension between 0 and 1 are simply shape sequences between shapes A
and B when the $t$-dimension value is fixed at zero.  If, however, the
$t$-dimension value is allowed to become positive as $s$ varies from 0 to 1,
then the intermediate shapes will take on some of the characteristics of
shape C.  In fact, the 5D implicit function actually captures an entire
family of shapes that are various blends between the three shapes.
Figure~\ref{fig:influence} illustrates some members of such a family of
shapes.

There is no reason to stop at three shapes.  It is possible to place four
shapes at the corners of a quadrilateral, five shapes around a pentagon, and
so on.  If we wish to use four shapes, then placing the constraints at the
corners of a quadrilateral using two additional dimensions would not allow
us to produce a shape that was arbitrary mixtures between the shapes.  In
order to do so, we can place the constraints in yet a higher dimension, in
effect placing the four shapes at the corners of a tetrahedron in $N+3$
dimensions, where $N$ is the dimension of the given shapes.

There are two related themes that guide our technique for shape
transformation.  The first is that shape transformation should be thought of
as a shape-creation problem in a higher dimension.  The second theme is that
better shape transformation sequences are produced when all of the problem
constraints are solved simultaneously--- in our case by using variational
interpolation.  Influence shapes are the result of taking these ideas to an
extreme.

\section{Conclusions and Future Work}
\label{sec:conclusions}

Our new approach uses variational interpolation to produce one implicit
function that describes an entire sequence of shapes.  Specific
characteristics of this approach include:

\begin{itemize}
  \item Smooth intermediate shapes
  \item Shape transformation in any number of dimensions
  \item Analytic solutions that are free of polygon and voxel artifacts
  \item Continuous surface normals for contour interpolation
  \item Contour slices may be at any orientation, even intersecting
\end{itemize}

This approach provides two new controls for creating shape transformation
sequences:

\begin{itemize}
  \item Separation distance gives local/global interpolation tradeoff
  \item May use influence shapes to control a transformation
\end{itemize}

The approach we have presented in this paper re-formulates the shape
interpolation problem as an interpolation problem in one higher dimension.
In essence, we treat the ``time'' dimension just like another spatial
dimension.  We have found that using the variational interpolation method
produces excellent results, but the mathematical literature abounds with
other interpolation methods.  An exciting avenue for future work is to
investigate what other interpolation techniques can also be used to create
implicit functions for shape transformation.  Another issue is whether shape
transformation methods can be made fast enough to allow a user interactive
control.  Finally, how might surface properties such as color and texture
be carried through intermediate objects?

\section{Acknowledgements}

This work owes a good deal to Andrew Glassner for getting us interested in
the shape transformation problem.  We thank our colleagues and the anonymous
reviewers for their helpful suggestions.  This work was funded by ONR grant
N00014-97-1-0223.


\bibliographystyle{plain}

\begin{thebibliography}{99}

\fontsize{8pt}{9pt}\selectfont
\itemsep 0in
\parskip 0in
\parsep 0in

\bibitem{Barequet96}
  Barequet, Gill, Daniel Shapiro and Ayellet Tal, ``History Consideration in
  Reconstructing Polyhedral Surfaces from Parallel Slices,'' {\em
  Proceedings of Visualization '96}, San Francisco, California, Oct. 27 --
  Nov. 1, 1996, pp. 149--156.

\bibitem{Barr84}
  Barr, Alan H., ``Global and Local Deformations of Solid Primitives,'' {\em
  Computer Graphics}, Vol. 18, No. 3 (SIGGRAPH 84), pp. 21--30.

\bibitem{Beier_Neely92}
  Beier, Thaddeus and Shawn Neely, ``Feature-Based Image Metamorphosis,''
  {\em Computer Graphics}, Vol. 26, No. 2 (SIGGRAPH 92), July 1992, pp.
  35--42.

\bibitem{Bloomenthal94}
  Bloomenthal, Jules, ``An Implicit Surface Polygonizer,'' in {\em Graphics
  Gems IV}, edited by Paul S. Heckbert, Academic Press, 1994, pp.  324--349.

\bibitem{Bookstein89}
  Bookstein, Fred L., ``Principal Warps: Thin Plate Splines and the
  Decomposition of Deformations,'' {\em IEEE Transactions on Pattern
  Analysis and Machine Intelligence}, Vol. 11, No. 6, June 1989, pp.
  567--585.

\bibitem{Celniker_Gossard91}
  Celniker, George and Dave Gossard, ``Deformable Curve and Surface
  Finite-Elements for Free-Form Shape Design,'' {\em Computer Graphics},
  Vol. 25, No. 4 (SIGGRAPH 91), July 1991, pp. 257--266.

\bibitem{Cohen-Or97}
  Cohen-Or, Daniel, David Levin and Amira Solomovici, ``Three Dimensional
  Distance Field Metamorphosis,'' {\em ACM Transactions on Graphics}, 1997.

\bibitem{Duchon77}
  Duchon, Jean, ``Splines Minimizing Rotation-Invariant Semi-Norms in
  Sobolev Spaces,'' in {\em Constructive Theory of Functions of Several
  Variables}, Lecture Notes in Mathematics, edited by A. Dolb and B. Eckmann,
  Springer-Verlag, 1977, pp. 85--100.

\bibitem{Duncan91}
  Duncan, Jody, ``A Once and Future War,'' {\em Cinefex}, No. 47 (entire
  issue devoted to the film Terminator 2), August 1991, pp. 4--59.

\bibitem{Fuchs77}
  Fuchs, H., Z. M. Kedem and S. P. Uselton, ``Optimal Surface Reconstruction
  from Planar Contours,'' {\em Communications of the ACM}, Vol. 20, No. 10,
  October 1977, pp. 693--702.

\bibitem{Golub_vanLoan96}
  Golub, Gene H. and Charles F. Ban Loan, {\em Matrix Computations}, John
  Hopkins University Press, 1996.

\bibitem{Gregory98}
  Gregory, Arthur, Andrei State, Ming C. Lin, Dinesh Manocha, Mark A.
  Livingston, ``Feature-based Surface Decomposition for Correspondence
  and Morphing between Polyhedra'', {\em Proceedings of Computer Animation},
  Philadelphia, PA., 1998.

\bibitem{He94}
  He, Taosong, Sidney Wang and Arie Kaufman, ``Wavelet- Based Volume
  Morphing,'' {\em Proceedings of Visualization '94}, Washington, D. C.,
  edited by Daniel Bergeron and Arie Kaufman, October 17-21, 1994, pp.
  85--92.

\bibitem{Herman92}
  Herman, Gabor T., Jingsheng Zheng and Carolyn A.   Bucholtz, ``Shape-Based
  Interpolation,'' {\em IEEE Computer Graphics and Applications}, Vol. 12,
  No. 3 (May 1992), pp. 69--79.

\bibitem{Hughes92}
  Hugues, John F., ``Scheduled Fourier Volume Morphing,'' {\em Computer
  Graphics}, Vol. 26, No. 2 (SIGGRAPH 92), July 1992, pp. 43--46.

\bibitem{Kaul_Rossignac91}
  Kaul, Anil and Jarek Rossignac, ``Solid- Interpolating Deformations:
  Construction and animation of PIPs,'' {\em Proceedings of Eurographics
  '91}, Vienna, Austria, 2-6 Sept. 1991, pp.   493--505.

\bibitem{Kent92}
  Kent, James R., Wayne E. Carlson and Richard E. Parent, ``Shape
  Transformation for Polyhedral Objects,'' {\em Computer Graphics}, Vol.
  26, No. 2 (SIGGRAPH 92), July 1992, pp. 47--54.

\bibitem{Lerios95}
  Lerios, Apostolos, Chase Garfinkle and Marc Levoy, ``Feature-Based Volume
  Metamorphosis,'' {\em Computer Graphics Proceedings}, Annual Conference
  Series (SIGGRAPH 95), pp. 449--456.

\bibitem{Levin86}
  Levin, David, ``Multidimensional Reconstruction by Set-valued
  Approximation,'' {\em IMA J. Numerical Analysis}, Vol. 6, 1986, pp.
  173--184.

\bibitem{Litwinowicz_Williams94}
  Litwinowicz, Peter and Lance Williams, ``Animating Images with Drawings,''
  {\em Computer Graphics Proceedings}, Annual Conference Series (SIGGRAPH
  94), pp. 409--412.

\bibitem{Lorenson_Cline87}
  Lorenson, William and Harvey E. Cline, ``Marching Cubes: A High Resolution
  3-D Surface Construction Algorithm,'' {\em Computer Graphics}, Vol. 21,
  No. 4 (SIGGRAPH 87), July 1987, pp. 163--169.

\bibitem{Meyers_Skinner91}
  Meyers, David and Shelley Skinner, ``Surfaces From Contours: The
  Correspondence and Branching Problems,'' {\em Proceedings of Graphics
  Interface '91}, Calgary, Alberta, 3-7 June 1991, pp. 246--254.

\bibitem{Payne_Toga92}
  Payne, Bradley A. and Arthur W. Toga, ``Distance Field Manipulation of
  Surface Models,'' {\em IEEE Computer Graphics and Applications}, Vol. 12,
  No. 1, January 1992, pp. 65--71.

\bibitem{Rossignac_Kaul94}
  Rossignac, Jarek and Anil Kaul, ``AGRELs and BIPs: Metamorphosis as a
  Bezier Curve in the Space of Polyhedra,'' {\em Proceedings of Eurographics
  '94}, Oslo, Norway, Sept. 12--16, 1994, pp.   179--184.

\bibitem{Sederberg_Greenwood92}
  Sederberg, Thomas W. and Eugene Greenwood, ``A Physically Based Approach
  to 2-D Shape Blending,'' {\em Computer Graphics}, Vol. 26, No. 2 (SIGGRAPH
  92), July 1992, pp. 25--34.

\bibitem{Sederberg_Parry86}
  Sederberg, Thomas W. and Scott R. Parry, ``Free-Form Deformations of
  Solid Geometric Models,'' {\em Computer Graphics}, Vol.   20, No. 4
  (SIGGRAPH 86), pp. 151--160.

\bibitem{Turk99}
  Turk, Greg and James F. O'Brien, ``Variational Implicit Surfaces,'' Tech
  Report GIT-GVU-99-15, Georgia Institute of Technology, May 1999, 9 pages.

\bibitem{Wolberg90}
  Wolberg, George, {\em Digital Image Warping}, IEEE Computer Society Press,
  Los Alamitos, California 1990.

\end{thebibliography}

\begin{figure}[!t]
\centerline{\epsfig{file=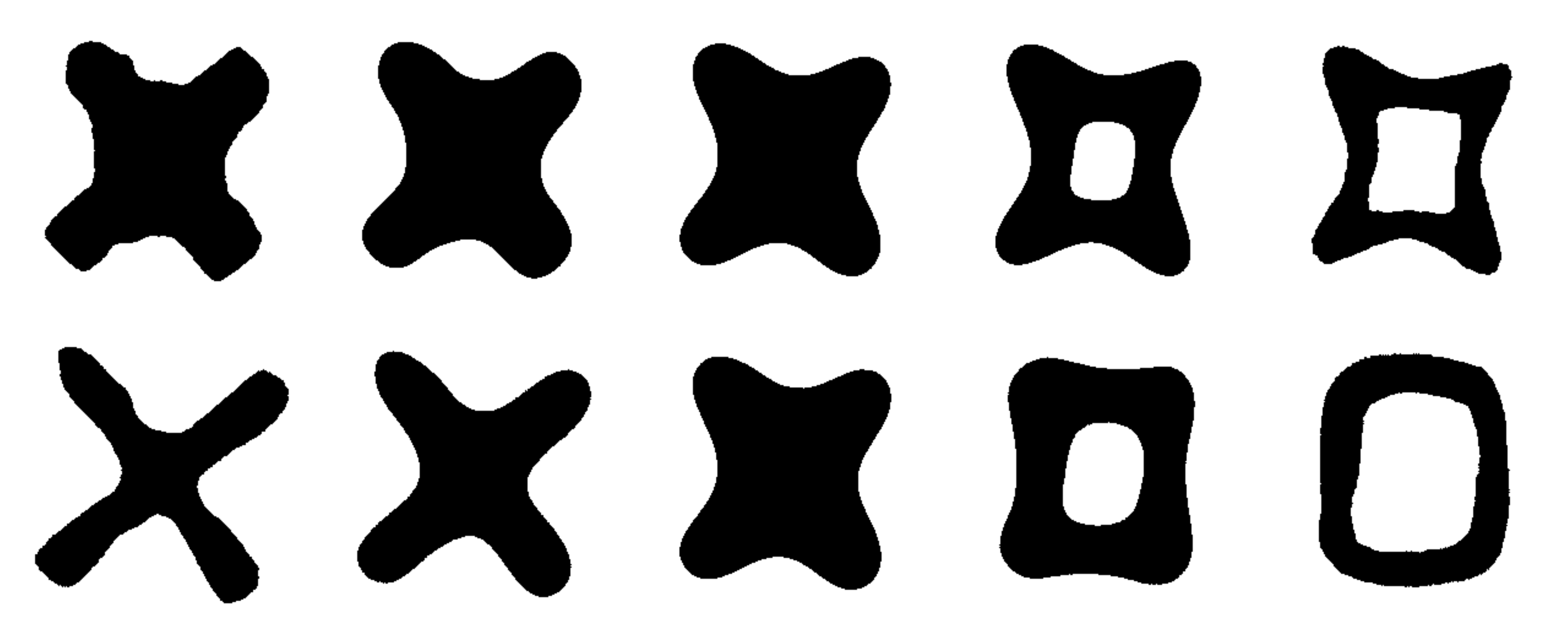,width=4.0in}}
\vspace*{-0.15in}
\caption{
  The extreme left and right shapes in the top row have been warped before
  creating the upper shape transformation sequence.  The lower row is
  an un-warped version of this sequence that gives the final transformation
  from an X to O.
}
\vspace*{-0.15in}
\label{fig:xo_warp_unwarp}
\end{figure}

\vspace*{-0.25in}

\section{Appendix: Warping}

Warping is a commonly used method of providing user control of shape
interpolation.  Although warping is not a focus of our research, for the
sake of completeness we describe below how warping may be used
together with our shape transformation technique.  Research on warping
(sometimes called deformation) include 
\cite{Barr84,Sederberg_Parry86,Wolberg90,Beier_Neely92,Lerios95,Cohen-Or97}.

For symmetry, we choose to warp each shape ``half-way'' to the other shape.
Given a set of user-supplied corresponding points between two shapes $A$ and
$B$, we construct two displacement warp functions ${\bf w}_A$ and ${\bf
w}_B$.  The function ${\bf w}_A$ specifies what values to add to locations
on shape $A$ in order to warp it half-way to shape $B$, and the warping
function ${\bf w}_B$ warps $B$ half of the way to $A$.

In what follows, we will describe the warping process for two-dimensional
shapes.  Let $\{{\bf a}_1, {\bf a}_2,\ldots, {\bf a}_k\}$ be a set of points
on shape $A$, and let $\{{\bf b}_1, {\bf b}_2,\ldots, {\bf b}_k\}$ be the
corresponding points on $B$.  We construct the two functions $w_A$ and $w_B$
such that $w_A({\bf a}_i) = ({\bf b}_i - {\bf a}_i)/{2}$ and $w_B({\bf b}_i)
= ({\bf a}_i - {\bf b}_i)/{2}$ hold for all $i$.  
Constructing these functions is another example of scattered data
interpolation which we can solve using variational techniques.  For
2D shapes, if ${\bf a}_i = (a^x_i, a^y_i)$ and ${\bf b}_i = (b^x_i, b^y_i)$,
then the $x$ component $w^x_A$of the displacement warp ${\bf w}_A$ has $k$
constraints at the positions ${\bf a}_i$ with values $(b^x_i - a^x_i) / 2$.
We invoke variational interpolation to satisfy these constraints, and do the
same to construct the $y$ component of the warp.  The function ${\bf w}_B$
is constructed similarly.  This is not a new technique, and researchers
who use thin-plate techniques to perform shape warping
include~\cite{Bookstein89,Litwinowicz_Williams94} and others.

In order to combine warping with shape transformation, we use these
functions to displace all of the boundary constraints of the given shapes.
These displaced boundary constraints are embedded in 3D (as described in
Section~\ref{sec:shape_interpolation}) and then variational interpolation is
used to create the implicit function that describes the entire shape
transformation sequence.  The result of this process is a three-dimensional
implicit function, each slice of which is an intermediate shape between two
warped shapes.  The top row of Figure~\ref{fig:xo_warp_unwarp} shows such
warped intermediate shapes.  We can think of the two ``ends'' of this
implicit function (at $t = 0$ and $t = t_{max}$) as being warped versions of
our original shapes.  In order to match the two original shapes, the surface
of this 3D implicit function needs to be unwarped.  To simplify the
equations, assume that the value of $t_{max}$ is 2.  If $t \le 1$ the
unwarping function $u(x,y,t)$ is:

\begin{equation}
 u(x,y,t) = (x + (1-t) w^x_A(x,y), y + (1-t) w^y_A(x,y), t)
\end{equation}

If $t > 1$ then the unwarping function is:

\begin{equation}
 u(x,y,t) = (x + (t-1) w^x_B(x,y), y + (t-1) w^y_B(x,y), t)
\end{equation}

At the extreme of $t = 0$, the warp $u(x,y,t)$ un-does the warping we used
for the first shape.  At $t = 2$, the function $u(x,y,t)$ reverses the
warping used for the second shape.  When $t = 1$ (the middle shape in the
sequence), no warp is performed.  The bottom sequence of shapes in
Figure~\ref{fig:xo_warp_unwarp} shows the result of the entire shape
transformation process that includes warping.  Both sequences in
Figure~\ref{fig:inf_the_end} were created using warping in addition to shape
transformation.

Although we have described the warping process for 2D shapes, the same
method may be used for shape transformation between 3D shapes.  For
Figure~\ref{fig:bunstar} (right), warping was used to align the bunny ears
to the points of the star.

\end{document}
\endinput


distill -resolution 200 -compresstext on -v -compatlevel 3.0 -colordownsample on -graydownsample on -monodownsample on -colordownsampletype average -graydownsampletype average -monodownsampletype average -colorres 200 -grayres 200 -monores 200 -colorcompr lzw -graycompr lzw -monocompr lzw -coloracs off -grayacs off -embedallfonts on schange.ps